# Kinetics of Hexagonal Cylinders to Face-centered Cubic Spheres Transition of Triblock Copolymer in Selective Solvent: Brownian Dynamics Simulation[*]

Minghai Li[†], Yongsheng Liu, Rama Bansil[‡]

*Department of Physics, Boston University, Boston, MA 02215, USA*

## Abstract

The kinetics of the transformation from the hexagonal packed cylinder (HEX) phase to the face-centered-cubic (FCC) phase was simulated using Brownian Dynamics for an ABA triblock copolymer in a selective solvent for the A block. The kinetics was obtained by instantaneously changing either the temperature of the system or the well-depth of the Lennard-Jones potential. Detailed analysis showed that the transformation occurred via a rippling mechanism. The simulation results indicated that the order-order transformation (OOT) was a nucleation and growth process when the temperature of the system instantly jumped from 0.8 to 0.5. The time evolution of the structure factor obtained by Fourier Transformation showed that the peak intensities of the HEX and FCC phases could be fit well by an Avrami equation.

## Introduction

It is well known that block copolymers exhibit a rich phase diagram with different ordered phases, such as 3 dimensional (3D) body center cubic (BCC)/

---

face center cubic (FCC) and the more complicated Gyroid and other bicontinuous phases, 2D hexagonal packed cylinder (HEX), and 1D lamellar (LAM) phases.[1-4] A variety of self assembled micellar domain shapes (spherical, cylindrical or planar sheets) can be obtained from a block copolymer by varying composition of and number of blocks, or by varying the polymer concentration, temperature and solvent selectivity in a block copolymer of fixed composition. Block copolymers, like lyotropic liquid crystals, offer a unique system to investigate transformations that simultaneously involve a change in the shape of the micellar domains and the symmetry of the underlying lattice, for example from HEX cylinder to BCC. While there are many studies of the equilibrium phase diagrams and thermodynamics of solvent mediated interactions in block copolymer systems, the kinetics is not so well understood. A few studies have been reported on the kinetics of the HEX cylinder to BCC sphere transition[5-8] but to the best of our knowledge, there is no published report on the kinetics on order-order transformation (OOT) of HEX cylinders to FCC spheres.

Computational simulation methods can provide the microscopic structural changes involved in the transformation between different phases. Several computational simulations using molecular dynamics (MD),[9-12] discrete MD,[13] Brownian Dynamics (BD),[14-19] Monte Carlo,[20-22] dissipative particle dynamics,[23] and time-dependent Ginsburg-Landau.[24] With this view we have undertaken a computational study on the kinetics of cylinders to spheres transition using Brownian Dynamics. Brownian Dynamics is particularly suited for simulating polymer solutions because it correctly models the Langevin dynamics for describing diffusion.[15] The solvent is treated implicitly. In the simulation, the system is coarse-grained such that the elemental unit is not a single molecule or even a single monomer of the polymer, but rather a sphere representing the center of the mass of a cluster of many molecules. This sphere (denoted as monomer or bead in the later text) moves according to Newton laws of motion. There are two time scales in the polymer solution system: the short time scale of the motion of the solvent molecules whose mass is much less than that of the coarse-grained polymeric monomer, and the long time scale of the motion of the polymeric monomers. Brownian dynamics only simulates the longer time scale of polymeric monomers and not the short time scale of the solvent motion. Thus compared to all atom molecular dynamics, BD is more efficient and saves computational time in simulating the polymer solution system. For example, BD methods have been used for simulating polymer flow,[16] phase diagram in surfactants modeled as sphere tethered to a chain[15,17] and in block copolymer melts,[14] solution,[18] and polymer brushes systems.[11,19] To the best of our knowledge, BD simulation has not been reported to study the kinetics of the HEX cylinders to FCC or BCC spheres transition for block copolymer in a selective solvent system.

Solvent selectivity further enriches the phase map and behavior of the block copolymers.[25] It is well known that in tri-blocks it is possible to obtain either isolated or bridged micelles depending on whether the solvent prefers the outer A block or the inner B block.[26] It is also possible to produce inverted micelles with the majority component in the cores by using a solvent selective for the minority block.[27] We investigated the equilibrium phase diagram of ABA triblocks in both A and B selective solvents[28] and observed isolated as well as bridged micelles. In this paper, we report Brownian Dynamics simulations to study the kinetics of the HEX cylinders to FCC spheres transition in triblock copolymer solution systems by either instantaneously quenching the temperature, $T$, or changing the well-depth of Lennard-Jones (L-J) potential, $\varepsilon$.

## Simulation Model

We simulated a symmetric triblock (ABA) copolymer in a selective solvent good for A block. The system consists of either 200 or 400 bead-spring chains in a cubic box. Each block of the copolymer chain has ten monomers of A or B, i.e., $A_{10}B_{10}A_{10}$, thus the total number of monomers in each triblock chain is 30 with a fraction of B monomers of 1/3. The A and B monomers are thermodynamically immiscible, thus the monomers of different types tend to separate from each other. A melt of such a triblock copolymer $A_{10}B_{10}A_{10}$ is similar in its behavior to the diblock $A_{10}B_5$ and thus expected to form the HEX cylinders structure, with the minority B monomers forming the cylindrical cores embedded in the matrix of the majority A monomers.[29]

The potential energy associated with the interaction between non-bonded monomers is given by either the standard Lennard-Jones (L-J) 6-12 potential or the modified version, the Weeks-Chandler-Andersen (WCA) potential, depending on the types of interacting monomers. Lennard-Jones (L-J) potential is used for the interaction between B-B monomers, which includes the attractive term and repulsive term,[30]

$$U_{L-J} = 4\varepsilon[(\frac{\sigma}{r})^{12} - (\frac{\sigma}{r})^{6}] - U_{L-J}(r_c), \quad r \leq r_c = 2.5\sigma$$
$$U_{L-J} = 0, \quad r > r_c \tag{1}$$

where σ is the diameter of a bead (set to 1 in simulation), $\varepsilon$ is the well-depth of L-J potential, and the cutoff radius $r_c = 2.5\sigma$. The aggregation of type B monomers is driven by non-zero well-depth, $\varepsilon$.

To describe the immiscibility between type A and B monomers, for A-B interactions, we use the repulsive Weeks-Chandler-Andersen (WCA) potential,[31]

$$U_{WCA} = 4\varepsilon[(\frac{\sigma}{r})^{12} - (\frac{\sigma}{r})^{6}] + \varepsilon, \quad r \leq r_c = 2^{1/6}\sigma$$
$$U_{WCA} = 0, \quad r > r_c \tag{2}$$

where the cutoff radius $r_c = 2^{1/6}\sigma$. Here the original L-J potential is truncated at its minimum $r_c = 2^{1/6}\sigma$ and shifted up to 0, so that it is always positive and purely repulsive. If the solvent is good to type A monomers, the WCA potential, instead of L-J potential, is used as type A-A interactions. Therefore, the degree of immiscibility between A and B monomers and the quality of solvent selectivity is determined by the parameter $\varepsilon/k_B T$,[15] where $T$ is temperature and $k_B$ is the Boltzmann's constant.

The interaction between two neighboring bonded beads on a polymer chain is modeled by a finitely extensible nonlinear elastic (FENE) potential,[12]

$$U_{FENE} = -15\varepsilon(\frac{R_0}{\sigma})^2 \ln[1 - (\frac{r}{R_0})^2], \quad r < R_0 = 1.5\sigma$$
$$U_{FENE} = \infty, \quad r \geq R_0, \tag{3}$$

where $R_0$ is maximum allowable separation between two neighboring beads on the same chains. The FENE potential together with L-J or WCA potential yields a minimum potential at the separation 0.97σ for the two neighboring bonded beads on a polymer chain.

In the BD simulation each bead is subjected to conservative forces $F_i^C$, friction forces $F_i^F$ and random forces $F_i^R$. The governing equation of motion[30] is $m\ddot{r}_i = F_i^C + F_i^F + F_i^R$, where $m$ is the mass of bead. The friction force is $F_i^F = -\gamma v_i = -6\pi\sigma\eta v_i$, where $\gamma$ is friction coefficient, $\eta$ is solvent viscosity and $v$ is velocity of the bead. The random force $F_i^R$ arises from the random bombardment of solvent molecules and the effect of the thermal heat bath on the individual bead. The random force and friction force acting as hot and cold sources, respectively, constitute a non-momentum conserving thermostat and obey the fluctuation dissipation theorem,[32]

$$<F_i^R(t)F_j^R(t')> = 6k_BT\gamma\delta_{ij}\delta(t-t'). \tag{4}$$

We applied periodic boundary conditions in all the simulations. A random disordered configuration with a bond length of 0.97σ for the neighboring beads on the same chain was generated as the initial configuration. The polymer volume fraction $\phi = NV_0/V$ was fixed during the simulation, where $N$ is total number of beads, $V_0 = \pi\sigma^3/6$ is the volume of an individual bead, and $V$ is the total system volume. The system size was chosen to be big enough to avoid finite size effects, yet not too large so as to obtain the equilibrium structure in a reasonable amount of computational time. The Verlet velocity method was applied and the time step $\Delta t$ = 0.01 with the time unit[14] $\sigma\sqrt{m/\varepsilon}$ was used to integrate the discretized equations of motion. Typically the equilibrium configurations were obtained after running for $10^6$ ~$10^7$ time steps. Several simulations with different initial configurations were performed to avoid being trapped in a local energy minimum state. Visual Molecular Dynamics (VMD)[33] software was used for visualization of the morphology.

## Results and Discussion

**Ordered Structures and Phase Map in Solvent Selective for the Outer Block**

We simulated the $A_{10}B_{10}A_{10}$ system at different temperatures for three different volume fractions $\phi =$ 20%, 25%, and 30% in cubic boxes. Because BD is a very time consuming technique, it is not suitable to simulate the entire phase diagram with this technique. The phase map of observed morphologies for the three concentrations simulated in this work is shown in Figure 1. All temperatures are given in units of $\varepsilon/k_B$. At $\phi=20\%$, we only observed FCC spheres at lower temperature T = 0.5 and disordered spheres at higher T = 1. At $\phi=25\%$, we observed FCC spheres at T < 0.6, and HEX cylinders at T > 0.6. At elevated temperatures, the cylinders bend and form worm-like micelles. At still higher temperatures, disordered micelles with ill-defined profiles are observed. At $\phi=30\%$, we observed HEX cylinders for T < 1.3, worm-like micelles at higher temperatures (T = 1.3), and disordered ill-defined micelles at still higher temperatures (T = 1.6). The snapshots of the HEX cylinders and FCC spheres at $\phi=25\%$ at T = 0.8 and 0.5, respectively, are shown in Figure 2. The A monomer is shown in red and B monomers in blue in all the snapshots displayed in this text.

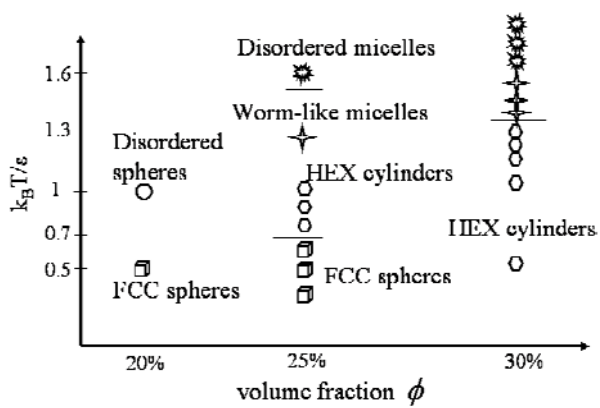

Figure 1. The phase map showing structures observed in the simulation for $A_{10}B_{10}A_{10}$ in an A-selective solvent at different volume fractions and temperatures.

As mentioned earlier, the melt of $A_{10}B_{10}A_{10}$ forms cylinders with B in the cores, which would transform to BCC spheres at high temperatures. For $A_{10}B_{10}A_{10}$ in solvent selective to outer A-block system, because solvent is poor to the minority B-block, the selectivity of the solvent further enhances the microphase separation tendency due to the incompatibility of A and B blocks. Thus we expect to obtain cylindrical or spherical micelles with the inner B blocks in the cores and the outer A blocks forming the solvated corona under certain concentrations and temperatures.

The existence of FCC lattice of spherical micelles in block copolymer melt has been predicted theoretically.[34,35] For example, using the self-consistent field theory, Matsen and Bates[36] predicted a close-packed spheres (CPS) packed into FCC or hexagonal close-packed (HCP) lattice to exist in a narrow region between BCC spheres and disordered phase in the phase diagram. However, such CPS morphology has not been identified experimentally in block copolymer melt so far. This may be due to the fact that the thermal fluctuations destabilize the long-range order of CPS and give rise to the disordered micelles.[37-39] Addition of selective low molecular weight solvent or homopolymer to the block copolymer gives rise to additional phase behaviors on the system. A stable FCC phase of spherical micelles has been reported in block copolymer solution systems[40-47] as well as in block copolymer/ homopolymer blends.[46] McConnell et al.[40] studied poly(styrene-*b*-isoprene) (SI) in decane, a selective solvent to PI, and observed the FCC and BCC spheres phase. They found the formation of FCC or BCC phase depends on the length of coronal layer thickness relative to the core radius, i.e., spherical micelles with thinner corona layer favors FCC phase due to the short-range inter-micellar repulsions, whereas for 'soft spheres' with thicker coronal layer BCC phase is favored. Hanley et al.[44] investigated the SI 20% in diethyl phthalate (DEP), a selective solvent to PS but poor for PI, and identified the FCC spheres phase at low temperature which transforms into HEX cylinders phase upon heating with the transition temperature of ~100 $^{\circ}$C. Recently, Park et al.[45] studied the SI in DEP solution system with different concentration and molecular weight and found the coexistence of FCC and HCP of spheres at low temperature. On increasing temperature, the system transforms from FCC/HCP spheres →BCC/HCP spheres →HEX cylinders.

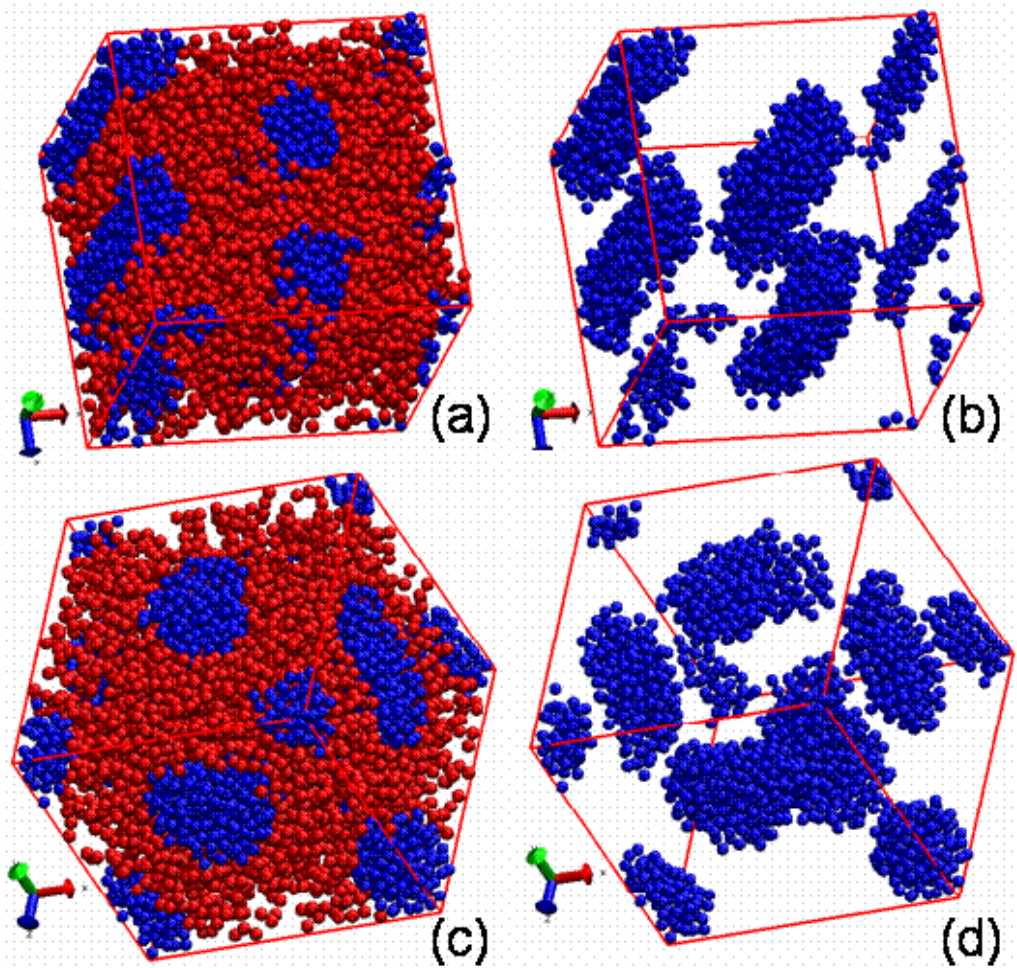

Figure 2. Snapshots of various ordered structures of $A_{10}B_{10}A_{10}$ in the solvent poor to B in a cubic box ($23.25\sigma \times 23.25\sigma \times 23.25\sigma$) at volume fraction $\phi = 25\%$. The monomer A is shown in red and B in blue. (a) HEX cylinders formed by B monomers embedded in a matrix of A monomers at $T = 0.8$ ($\varepsilon/k_B$); (b) For clarity, the same snapshot as (a) is shown with only B monomers displayed. (c) FCC spheres formed by B monomers embedded in A matrix at $T = 0.5$ ($\varepsilon/k_B$). (d) The same snapshot as (c) is shown with only B monomers displayed for clarity.

McConnell et al.[40] found that the formation of lattice structure of spherical micelles depends on $\xi = \langle L \rangle_h / R_c$, the ratio of the coronal layer thickness ($\langle L \rangle_h$) to the radius of the core ($R_c$). According to the phase diagram, as shown in Figure 3 in ref. 40, for ξ < 1.5, the system favors the formation of FCC lattice, whereas for the ξ > 1.5, the formation of BCC lattice is favored. We found that in our simulation at $\phi = 25\%$ and T = 0.5, the radius of the sphere is ~7σ and nearest-neighbor distance is ~16.5σ, thus ξ ≈ 1.4, assuming the coronal layer thickness is obtained by subtracting the core radius from the nearest-neighbor distance. This ratio ξ ≈ 1.4 lies in the region of FCC in the phase diagram, therefore the observation of FCC lattice in our simulation is exactly as predicted.

**Kinetics of HEX to FCC Transition**

**Time Evolution of Structure Following Temperature Quench**

We investigated the kinetics of the transition from HEX cylinders to FCC spheres for the system of $A_{10}B_{10}A_{10}$ in the A-selective solvent in a rectangular box with size of $23.25\sigma \times 23.25\sigma \times 46.50\sigma$ at the volume fraction 25% by instantaneously quenching the temperature from 0.8 to 0.5, with a fixed well-depth of L-J, $\varepsilon = 1$.

At T = 0.8, the system is in the HEX cylinder ordered structure (Figure 2(a)). Following the quench we run the simulation at low T = 0.5 until the FCC structure is obtained. The time evolution of the developing morphology for HEX cylinders to FCC spheres transition is shown in Figure 3.

At time t = 0, corresponding to the instantaneous quench time, the system was in the HEX cylinder phase. At t = $10^5$ time steps at least one cylinder shows a ripple, indicating the beginning of the breakup into spheres. At t = $4*10^5$ time steps, the middle cylinder is completely broken into the spheres. This induces the neighboring cylinders to ripple. The neighboring cylinders then break up and form spheres at different positions along the cylinder axis. At t = 29*10$^5$ time steps, the system shows a FCC spheres structure. It appears that after the cylinders broke up, the spheres rearranged to the FCC lattice.

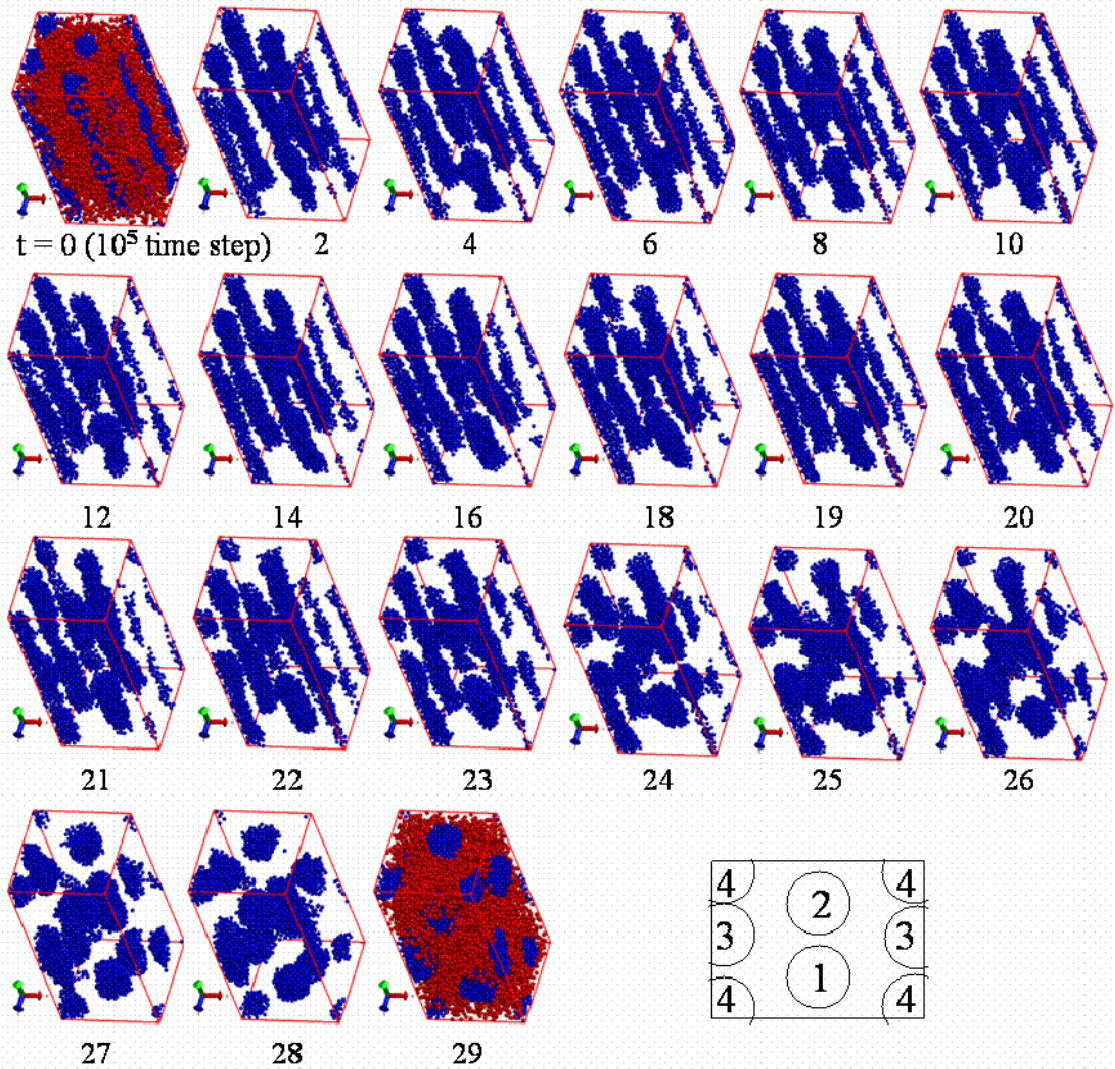

Figure 3: Time sequence of morphology of $A_{10}B_{10}A_{10}$ for HEX cylinders to FCC transition following a temperature jump from 0.8 to 0.5 (with $\varepsilon$ = 1) in a rectangular box with the size of $23.25\sigma \times 23.25\sigma \times 46.50\sigma$ at volume fraction $\phi = 25\%$. Time is indicated in the scale of $10^5$ time steps. The right bottom insert shows the index numbers of all four cylinders of the initial configuration from the top view. A is removed for clarity for some of the snapshots.

Simulations with longer cylinders were carried out with initial configuration generated by repeating the box of Figure 2(a) twice along the cylinder axis dimension, thereby doubling the length of the cylinders. The time sequence of the morphology during the transition is shown in Figure 3. Each cylinder of the initial configuration was indexed with a number shown in the insert of Figure 3.

The results shown in Figure 3 are consistent with nucleation and growth mechanism of the transition from cylinders to spheres. In the nucleation and growth scenario, the cylinders are meta-stable with respect to the modulation, thus some parts of the cylinders would develop ripples while others would remain intact, and the front of the modulation would advance with time along the cylinder axis, as described by Matsen.[48] The ripples would induce rippling in the neighboring cylinders. In contrast, in the spinodal decomposition scenario, the cylinders are unstable with respect to the modulation, and thus ripples form over the entire length of the cylinders. Moreover, if the modulations are correlated with neighboring cylinders, the epitaxial transition is possible.

To see more clearly how one cylinder transforms to spheres, the first cylinder is extracted and the time sequence of the snapshots of its profile is shown in Figure 4. Figure 4 shows that at t = 2*10$^5$ time steps the cylinder began to pinch at its middle and it broke at the middle at 3*10$^5$ time steps. Small fluctuations of the rest of the cylinder were present until t = 22*10$^5$ time step, when the ripples on the rest of the cylinder began to grow. The cylinder eventually broke up into two ellipsoidal clusters at t = 23*10$^5$ time steps. The two clusters became spherical at t = 29*10$^5$ time steps.

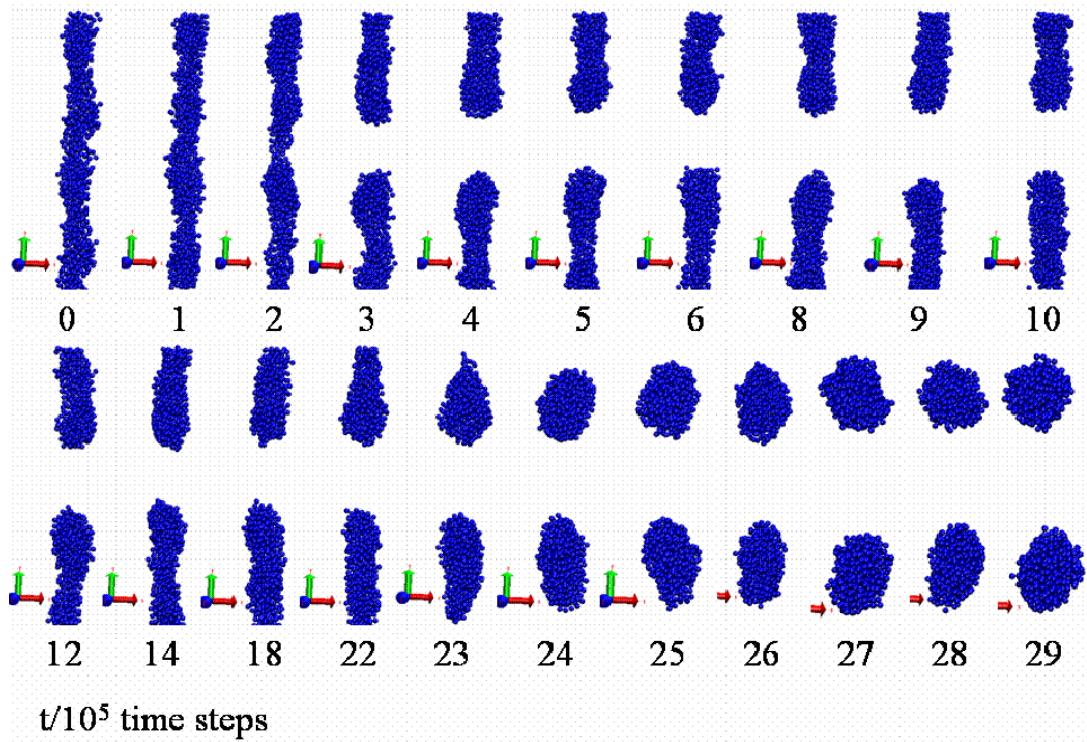

Figure 4. Time sequence of snapshots for the profile of the 1$^{st}$ cylinder extracted from Figure 3. Time is indicated in the scale of 10$^5$ time steps.

The simulation results described above are qualitatively in agreement with our previous paper[8] supporting the ripple mechanism for the HEX cylinders to BCC spheres phase transition in poly(styrene-*b*-ethylene-*co*-butylene-*b*-styrene) in mineral oil, a selective solvent for the middle block, using time-resolved SAXS following various temperature jumps. At a concentration of 45%, this system exhibits a HEX cylinder phase at lower temperature and undergoes an order-order transition (OOT) from HEX cylinders to BCC spheres upon heating. We observed that the transition occurred via a nucleation and growth mechanism for a shallow temperature jump and a spinodal decomposition mechanism with continuous ordering for a deep temperature jump. The scattering data was found to be in good agreement with scattering profiles calculated using a model rippled cylinder form factor and the phases between adjacent cylinders chosen to satisfy the epitaxial relationship.[8]

**Density Profiles**

To quantitatively study this process, we calculate the density profile of each cylinder in the following way: all the monomers that are poor to solvent belonging to a certain cylinder are extracted out; then the monomers are binned into different sections according to the position of center of each monomer along the cylinder axis (y axis in this case), and the number of monomers falling into each section is taken as the density profile vs. the section number, as plotted in Figure 5. One can see from Figure 5(a) that cylinder 1 breaks up at y ~ 25σ at t =3*10$^5$ time steps; at the same time two bulges appear and begin to develop at y ~ 17σ and 37σ, respectively. The bulges grow and become better defined as time increases. We note that the center of bulges moves and the fronts of the fluctuation advance to two sides of cylinder. At around t = 22~23*10$^5$ time steps the pinch or depletion appears and begins to grow around the vicinity of the ends of the cylinder. This observation is consistent with the snapshots displayed in the Figure 4. From the density profile of cylinder 2, shown in Figure 5(b), one can see that the cylinder 2 begins to fluctuate at $5*10^5$ time steps, while the fluctuation does not grow until t = $23*10^5$ time steps, when the bulge begins to grow at y ~ 22σ. One of the depletions appears at $24*10^5$ time step at y ~ 12σ and the other at $27*10^5$ time step at y ~ 35σ; The density profiles of cylinder 3 and 4 which are not shown here have the similar time evolution behaviors to cylinder 1 and 2, respectively. It also can be seen from the density profile that the four cylinders ripple and break up in a cooperative way at the different positions along the cylinder axis. All observations are in good agreement with scenario depicted in the nucleation and growth mechanism.

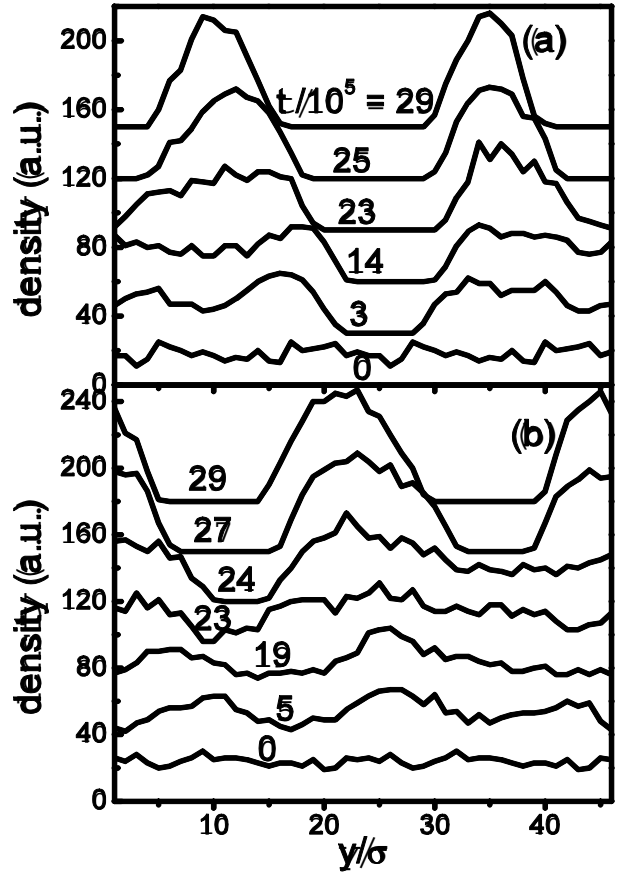

Figure 5. Time evolution of density profiles of two cylinders following a temperature quench from 0.8 to 0.5: (a) for cylinder 1; (b) for cylinder 2. The data is shifted vertically for clarity. Time is indicated in the scale of $10^5$ time steps.

**Fourier Transform of Pair Density Distribution**

The Fourier transform of the pair density distribution, i.e., the scattering intensities of configurations are calculated in two steps. First, the structure factor is given by $S(\vec{q}) = |\sum_{j \in B}(-i\vec{q}\cdot\vec{r}_j)|^2$, where $\vec{q}$ is scattering vector and $\vec{r}_j$ is position vector of B-type monomer. The sum is made only over all B-type monomers and the scattering contributed from A-type monomers is regarded as the background. Second, the azimuthally average scattering intensity $I(q)$ is calculated by numerical integration of $S(\vec{q})$ over the angular space as $I(q) = \int S(\vec{q}) \cdot d\Omega/4\pi$, where $d\Omega$ is the element of solid angle. The results are shown in Figure 6. The first three peaks at t = 0 are in good agreement with HEX predictions (with relative peak positions $1:\sqrt{3}:\sqrt{4}$); the first four peak at t = 29*10$^5$ time steps are also in good agreement with FCC predictions (with relative peak positions $1:\sqrt{4/3}:\sqrt{8/3}:\sqrt{11/3}$). From Figure 6, one can see that as time increases, small changes in scattering intensity are present until t = 24*10$^5$ time steps, when a new peak emerges and grows in the primary peak region, indicating a dramatic change in the transition process. This is consistent with the observation from the snapshot. As shown in Figure 3, at t = 24*10$^5$ time steps all four cylinders break, and some spheres are about to appear at a FCC lattice.

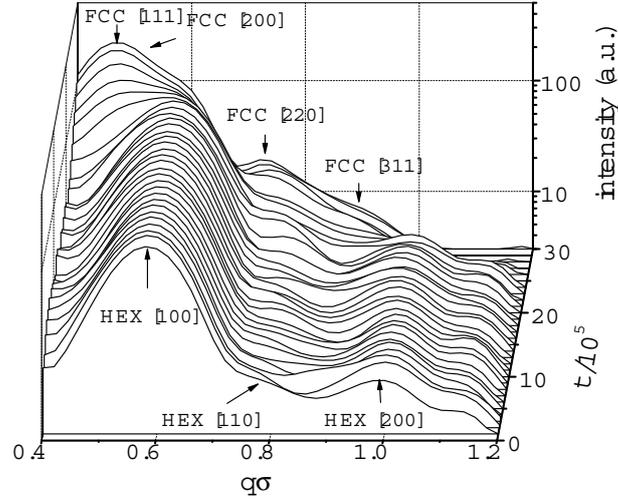

Figure 6. Time evolution of Fourier transform of the pair density distribution following a temperature quench from 0.8 to 0.5. The data is shifted vertically for clarity. Time is indicated in the scale of $10^5$ time steps. The first few relative peak positions for HEX and FCC are marked.

For primary peak, one can see two primary peaks coexisting: $I_{[100],HEX}$, i.e., HEX [100] Bragg peak and $I_{[111],FCC}$ for FCC [111] Bragg peak. They are distinct because their positions are apart far enough. The time evolutions of $I_{[100],HEX}$ and $I_{[111],FCC}$ are plotted in Figure 7. Avrami equation is often used to describe the nucleation and growth process. As expected, $I_{[100],HEX}$ decreases and $I_{[111],FCC}$ increases in a stretched exponential way, both of which can be well fitted by the Avrami equation:

$$I(t) - I(t_0) = (I(t_\infty) - I(t_0))(1 - e^{-((t-t_0)/\tau)^n}). \qquad (5)$$

The fitting results are also shown in Figure 7. The Avrami parameter n for HEX and FCC are 2.5 and 3.2, respectively, which agree with the fitting result n=3 in ref. 49.

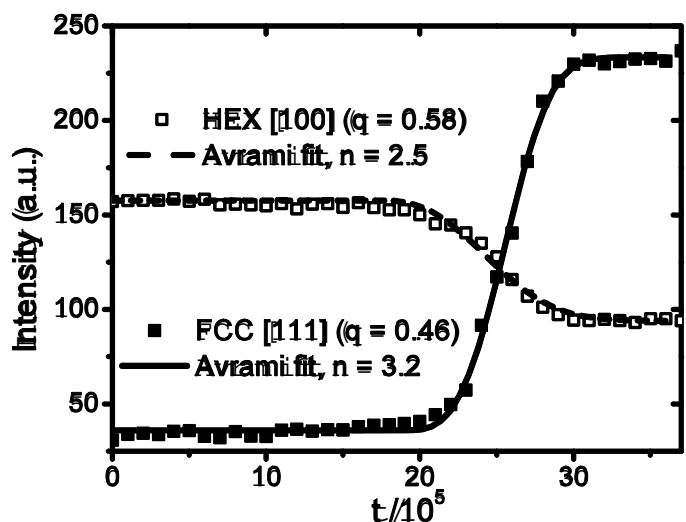

Figure 7. Time evolution of the intensities of primary peaks for HEX and FCC, following a temperature quench from 0.8 to 0.5.  The results of Avrami fitting are shown with lines.

Nie[49] reported a study of the FCC spheres to HEX cylinders transition in SI 40% w/v in dimethyl phthalate, a selective solvent to polystyrene. This diblock solution system exhibits an OOT from FCC spheres to HEX cylinders upon heating with transition temperature of 110 $^{o}$C. Two stages were observed in the transition. The early stage was described well by an Avrami equation,[50]  i.e. a stretched exponential growth with an exponent n = 3 or 4, indicating the growth of HEX from FCC involves a 2 or 3 dimensional mechanism.[51] The later stage was found similar to the secondary crystallization process observed in the kinetics of crystallization in homopolymers.[52]

The values of n we obtained in simulation are different from that obtained in the SAXS data for the HEX to BCC transition[8] and the FCC to BCC transition[53] (both with the typical value 1~1.5), indicating some differences in the growth mechanism. The overall behavior of the primary peaks agree well with the first stage reported by Nie[49] but no second stage is observed because the system simulated here is too small to observe the growth and coarsening of FCC micro-domains, as observed in the experiments.

**Simulations Following Jumps of Well-depth of L-J Potential.**

There is more than one way to generate a transformation by jumping either the temperature or L-J potential well-depth. An alternative approach to investigate the kinetics of HEX cylinders to FCC transition is to instantaneously jump the well-depth of L-J potential, $\varepsilon$. During the simulation the value of $T$ is fixed so that the jump does not change the average speed of monomers, and thus it saves the computation time. Indeed, the effect on thermodynamic behavior of increasing the value of $\varepsilon$ is equivalent to decreasing the value of $T$ because the temperature is measured in the scale of $\varepsilon/k_B$.

We started from the initial configuration of HEX cylinders for $A_{10}B_{10}A_{10}$ at $T$ = 0.8, $\phi = 25\%$, $\varepsilon$ =1, as in the temperature jump study. We run the simulation following an instantaneous jump of $\varepsilon$ from 1 to various higher values, $\varepsilon_f$ = 1.25~10, while fixing value of $T$ = 0.8. The snapshots of the configurations for all $\varepsilon$-jumps after running for certain time steps are shown in Figure 8.

For $\varepsilon$ jump to $\varepsilon_f$ = 1.25, as shown in Figure 8(a), one cylinder broke up and formed spheres and the other formed ripples at t = $10^7$ time steps. Thus, it was still in the intermediate stage at this time and undergoing further transition. For $\varepsilon$ jump to $\varepsilon_f$ = 1.5, as shown in Figure 8(b), the system formed the FCC at t = $3*10^6$ time steps. In fact, it already formed the FCC at earlier time. For $\varepsilon$ jump to $\varepsilon_f$ = 1.75, at t = $10^7$ time step as shown in Figure 8(c), the cylinders are broken. Some of these broken cylinders formed spheres, and others formed the short cylinders which may change into spheres at a later time. In contrast to these cases, for $\varepsilon$ jump to $\varepsilon_f$ = 2, at t = $10^7$ time steps as shown in Figure 8(d), the cylinders broke up to form short cylinders. These short cylinders persisted until a later time. For $\varepsilon$ jump to $\varepsilon_f$ = 2.5 and 3, at t = $10^7$ time steps as shown in Figure 8(e) and (f), respectively, the systems show behaviors similar to that of the $\varepsilon$ jump to 2. In these cases, however, sizes of resulting cylinders are a little longer. For $\varepsilon$ jump to $\varepsilon_f$ = 4, as shown in Figure 8(g), the cylinders are still all intact even at t = $10^7$ time steps and remain this way at later times. For higher $\varepsilon$ jumps to $\varepsilon_f$ = 8 and 10, the behavior is similar to the jump of $\varepsilon$ to 4. These systems are shown in Figure 8(h) and (i) at t = $5*10^6$ and $10^7$ time steps, respectively. The simulation time step is chosen as 0.005 for these two systems.

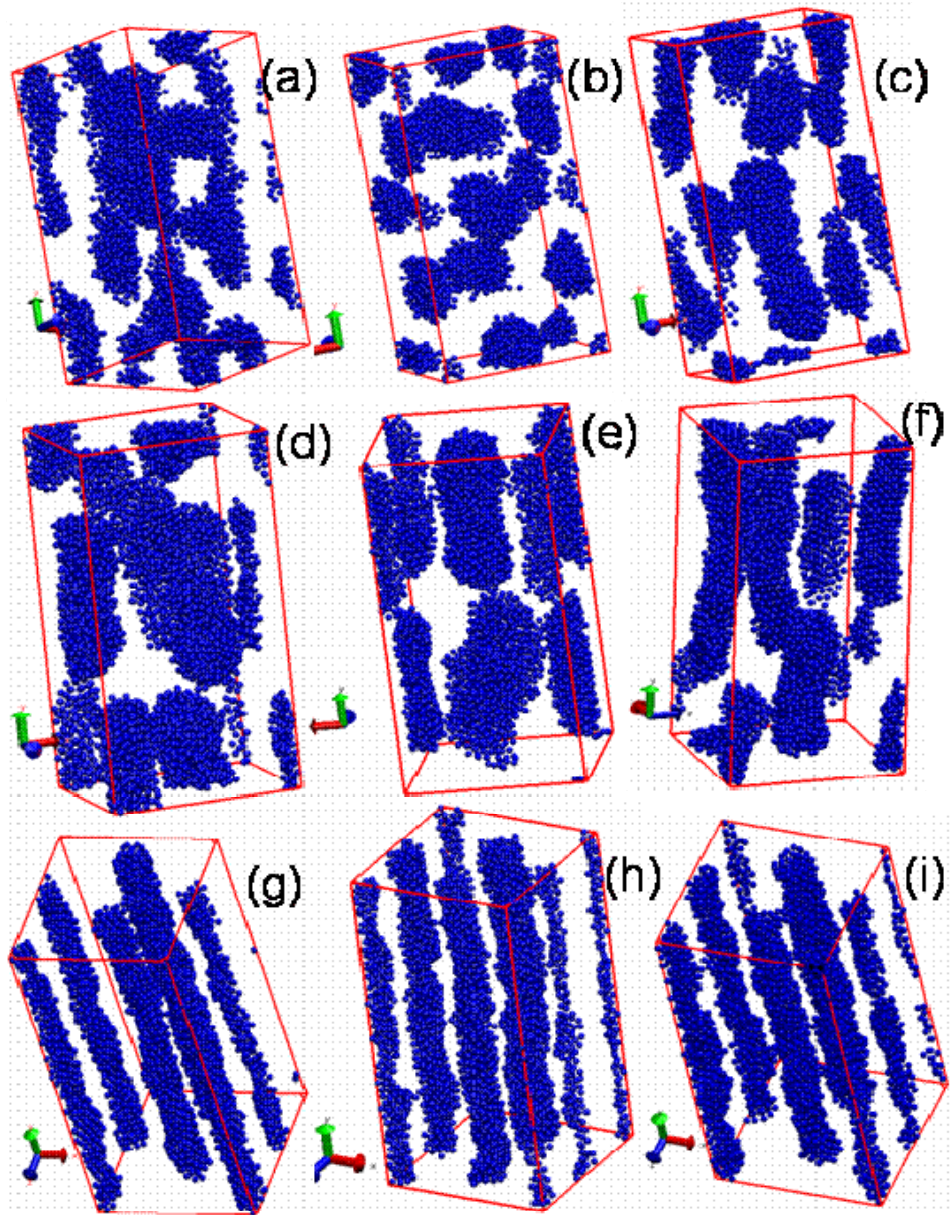

Figure 8. The snapshots of various $\varepsilon$ jumps for $A_{10}B_{10}A_{10}$ at $T = 0.8$, $\phi = 25\%$. All initial $\varepsilon = 1$. (a) $\varepsilon$ jump to 1.25, at t = $10^7$ time step; (b) $\varepsilon$ jump to 1.5, at t = $3*10^6$ time step; (c) $\varepsilon$ jump to 1.75, at t = $10^7$ time step; (d) $\varepsilon$ jump to 2, at t = $9.4*10^6$ time step; (e) $\varepsilon$ jump to 2.5, at t = $10^7$ time step; (f) $\varepsilon$ jump to 3, at t = $10^7$ time step; (g) $\varepsilon$ jump to 4, at t = $10^7$ time step; (h) $\varepsilon$ jump to 8, at t = $5*10^6$ step; (i) $\varepsilon$ jump to 10, at t = $10^7$ step. For all $\varepsilon$ jumps, the value of $T$ is fixed.

From the snapshots for the various $\varepsilon$ jumps simulated in this work and shown in Figure 8, one can see the existence of an optimum $\varepsilon$ jump to $\varepsilon_f \approx 1.5$, where the transition of HEX cylinders to FCC spheres occurs quickly via the nucleation and growth mechanism. For $\varepsilon$ jump to a value below this optimum, such as $\varepsilon_f$ = 1.25, the transition occurs slowly. For $\varepsilon$ jump to a value above 1.5 but less than 2, such as $\varepsilon_f$ = 1.75, the transition occurs slowly and it may take a long time to complete the transition. For $2 \leq \varepsilon_f \leq 3$, the cylinders break up but form short cylinders instead of spheres, and the transition may not be complete within our computational time limit. For $\varepsilon_f \geq 4$, the cylinders do not break up within our computational time limit, and the structure is frozen due to a kinetic trap.

**Conclusion**

In this paper, we report Brownian Dynamics simulation to study the kinetics of HEX cylinders to FCC spheres transition in triblock copolymer $A_{10}B_{10}A_{10}$ in an A-selective solvent following temperature quenches or well-depth of L-J jumps. We first observe the HEX cylinders and FCC spheres at temperature T = 0.8 and 0.5($\varepsilon/k_B$), respectively, at the volume fraction $\phi = 25\%$. We then use the HEX cylinders as initial configuration and quench the temperature from 0.8 to 0.5 while fixing $\varepsilon$ to study the kinetics of the transition from HEX cylinders to FCC spheres. The snapshots and density profiles show that one cylinder breaks at its middle part at early time and the front of the ripple begin to advance to the rest of the cylinder. It takes a long time for this cylinder to break up into two spheres. During this time, the ripple induces the formation and development of ripples of the neighboring cylinders. The Fourier transform of pair density distribution is performed and it indicates a dramatic change at t = 24*10$^5$ time steps, which is consistent with the observation from the snapshots that at this moment all the cylinders break up and the spheres are about to appear on the FCC lattice. The observation agrees well with the nucleation and growth mechanism.

We study the kinetics of this transition following $\varepsilon$ jump from 1 to various higher value while fixing the value of T = 0.8. The study shows an optimum $\varepsilon$ jump to $\varepsilon_f \approx 1.5$, under which the transition occurs easily. For $\varepsilon_f$ <1.5, the transition is completed in a longer time. For 1.5 < $\varepsilon_f$ < 2, the transition occurs much more slowly but could be completed within our computational time limit. For 2 ≤ $\varepsilon_f$ ≤ 3, the cylinders break up into short cylinders, and the transition may not be complete within our computational time limit. For $\varepsilon_f$ ≥ 4, the cylinders keep intact within our computational time limit, because the structure is frozen and kinetics is trapped.

We also studied the triblock copolymers $A_5B_{20}A_5$ and $A_{10}B_{10}A_{10}$ in a solvent selective to the inner B block. For the former system at $\phi = 35\%$, HEX cylinders and FCC spheres ordered structures are observed at T = 0.7 and 0.5, respectively. For the latter system at $\phi = 30\%$, LAM and HPL ordered structures are obtained at T = 1 and 0.5, respectively.

**Acknowledgements.** This research was supported by the National Science Foundation, Division of Materials Research (Grant No. NSF-DMR 0804784). We acknowledge the support of Boston University's Scientific Computation and Visualization group also supported by NSF for computational resources. We thank Prof. Sharon Glotzer, Prof. Bill Klein, Prof. Chi Wu, Dr. Zhenli Zhang, Dr. Zhenhua Wu, Dr. Rachele Dominguez, and Dr. Kipton Barros for helpful discussions.